\documentclass{edp-conf}
\usepackage{graphicx}

\newcommand{\ha}{H$\alpha$}
\newcommand{\hi}{\mbox{H{\sc i}}}

\begin{document}

\TitreGlobal{SF2A 2004}

%%-----------------------------
%%      the top matter
%%-----------------------------
\title{\vspace*{-1.6cm} A Virgo  high-resolution \ha\ kinematical survey}
\author{Chemin, L.$^{1,}$}
\address{D\'epartement de Physique, Universit\'e de Montr\'eal, C.P. 6128, Succ. centre-ville, Montr\'eal, Qc, Canada, H3C 3J7}
\address{Observatoire de Paris, section Meudon, GEPI, CNRS UMR 8111 \& Universit\'e Paris 7, 5 Pl. Janssen, 92195 Meudon, France}
\author{Balkowski, C.$^{2}$}
\author{Cayatte, V.$^{2}$}
\author{Adami, C.}\address{Laboratoire d'Astrophysique de Marseille, 2 Pl. Le Verrier, 13248 Marseille, France}
\author{Amram, P.$^{3}$}
\author{Boselli, A.$^{3}$}
\author{Boulesteix, J.$^{3}$} 
\author{Carignan, C.$^{1}$}
\author{Garrido, O.$^{2,3}$}
\author{Hernandez, O.$^{1,3}$}
\author{Marcelin, M.$^{3}$}
\author{Vollmer, B.}\address{Observatoire Astronomique de Strasbourg, CDS, UMR 7550, 11 rue de l'Universit\'e, 67000 Strasbourg, France}
\runningtitle{A Virgo \ha\ survey}
\setcounter{page}{237}
% Keep this line, even if the page will be settled afterwards..
\index{Chemin, L.}
\index{Balkowski, C.}
\index{Cayatte, V.}
\index{Adami, C.}
\index{Amram, P.}
\index{Boselli, A.}
\index{Boulesteix, J.}
\index{Carignan, C.}
\index{Garrido, O.}
\index{Hernandez, O.}
\index{Marcelin, M.}
\index{Vollmer, B.}
% Repeat the authors here, this will help to make the final index

\maketitle
%\begin{abstract} 
%\end{abstract}
%
%%-----------------------------
%%      your text
%%-----------------------------
\section{Abstract}\vspace*{-0.4cm}
 We have completed a survey of 30 Virgo cluster galaxies in the \ha\ emission-line 
using Fabry-Perot interferometry. The goal of the survey is to obtain a high angular resolution sample 
of velocity fields of spirals and to study the environmental effects on their kinematics and dynamics. 
\vspace*{-0.4cm}
\section{Global result and scientific goals}
\begin{figure}
\begin{center}
  \includegraphics[width=12.0cm]{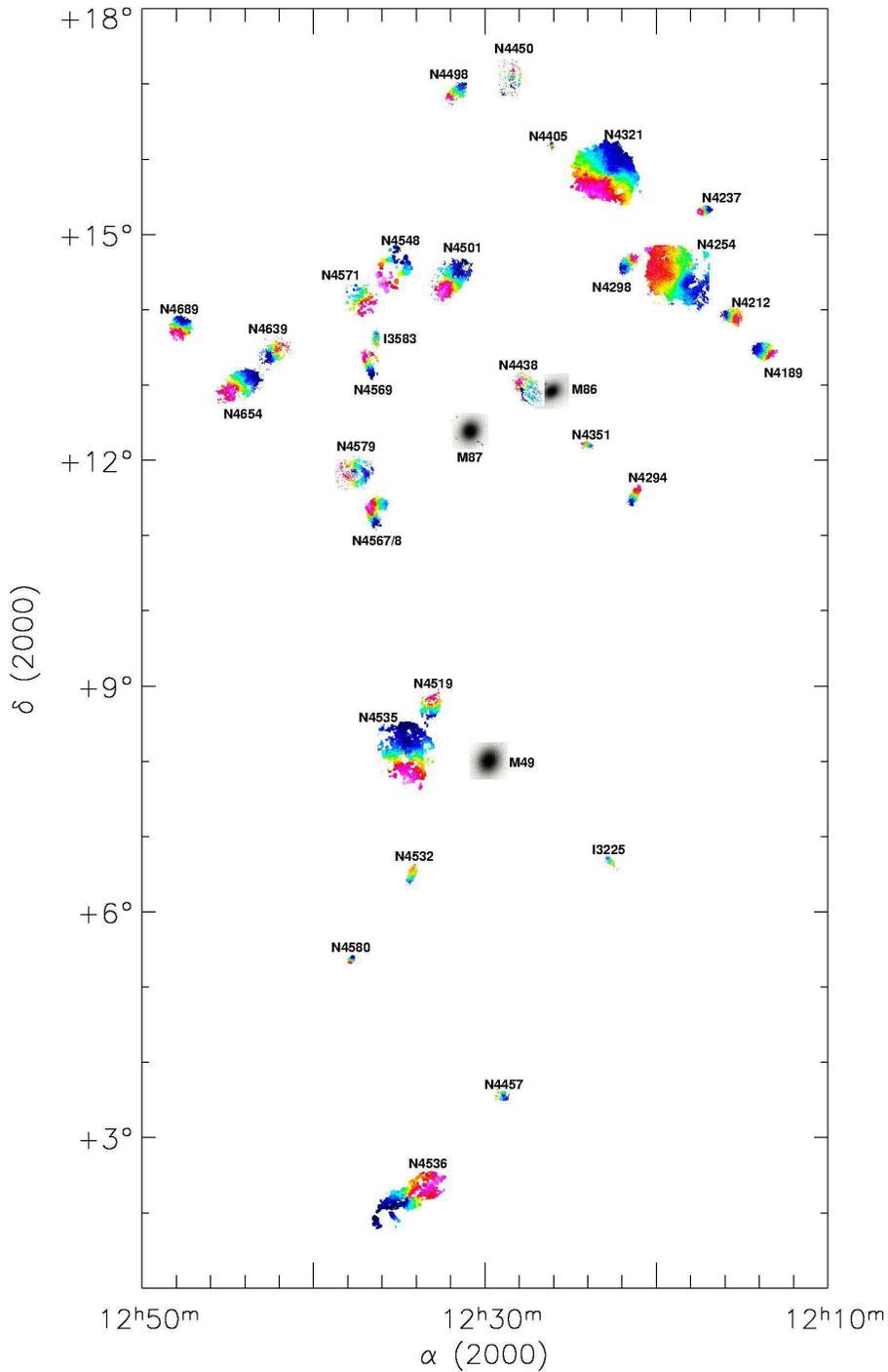}
\end{center}
\caption[]{A view of the Virgo cluster galaxy velocity fields map. For clarity reasons, the angular scale of each of the 30 galaxies has been enlarged by a factor of $\sim 9.5$ w.r.t. the real
scale. 
The locations of the 3 large ellipticals M49, M86 and M87 are shown using  broad-band optical images. The colour scale is blue-darker shades (red-lighter shades) for the approaching (receding respectively) 
side of galaxies. The velocity range of each galaxy will be presented in Chemin et al. (in prep.).}
\label{figposvf}
\end{figure}

\vspace*{-0.4cm}We present here an overall view of the 30  velocity fields in the Virgo cluster (Fig.~\ref{figposvf}). 
The whole data mining procedure and FP-\ha\  catalog will be fully  described in Chemin et al. (2004, to be submitted). 
 Previous results for a sub-sample of 9 galaxies were presented in Chemin (2003) and for the strongly perturbed spiral galaxy NGC 4438 in Chemin et al. (2004, submitted).
  Forthcoming papers  will detail the analyses and results  for the whole sample or for individual galaxies:   
 comparison with the \hi\ and CO   (Cayatte et al. 1990, Sofue et al. 2003) and long-slit (Rubin et al. 1999) Virgo catalogs, 
 tilted-ring models of velocity fields, rotation curves, residual velocity fields, 
 determination of  bar pattern-speeds for barred galaxies, dark matter distribution and also comparison with isolated galaxies (Garrido et al. 2002). 
 The high-resolution data should also be used to compare with results from numerical simulations 
 of ram-pressure stripping and/or tidal interactions (Vollmer et al. 2001).\\
 %%-----------------------------
%%      your bibliography
%%-----------------------------

\end{document}